# Heavy Ion Beams for Study of Thermophysical Properties

("HIB-2000" *Moscow, ITEP, May 29-31*)

## *Iosilevski I.L.*

(Moscow Institute of Physics and Technology)

= = = = = = = = = = = = = = = = = = = = = =

*Milestones*
*Priorities and preferences*
*Physical phenomena*
*Problem of uniform heating*
*Heating of highly dispersed porous materials*
*Presumed pressure−temperature trajectories*
*Measurement of equilibrium thermal expansion*
*Appendix A: "An Advanced Equation of State of $UO_2$*
  *up to the Critical Point*

\*\*\*\*\*\*\*\*\*\*\*\*\*\*\*\*\*\*\*\*\*\*\*\*\*\*\*\*\*\*\*\*\*\*\*\*\*\*\*\*\*\*\*\*\*\*\*\*\*\*

## MILESTONES

**We should change essentially our priorities in HIB development when we want to use the HIB for study of thermophysical properties of materials.**

**It seems to be realistic that we could obtain <u>*quasi-isobaric*</u> regime of heating if we combine the HIB energy deposition with the use of highly dispersed <u>*porous*</u> material as an irradiating sample.**

**When heating the <u>*evaporating porous*</u> material we could obtain the regime that is close to <u>*saturation curve*</u> in pressure-enthalpy coordinates.**

**In frames of this technique HIB could became an uncompetitive tool for study of <u>*phase transition*</u> phenomenon for a wide number of materials with high-temperature location of <u>*critical point*</u>.**

**Evaporation in <u>*Uranium*</u> is one of the most tempting candidates to be studied under HIB heating in such manner. When being successful this experiment has a good chance to resolve the old contradiction within the problem of <u>*Uranium critical point parameters*</u> estimations.**

**The heating by HIB seems to be especially promising as an effective tool for systematic study of so-called <u>*non-congruent*</u> phase transition – striking and mostly unusual sort of high-temperature phase equilibrium in chemically active strongly coupled plasmas. <u>*Phase transition*</u> in <u>*Uranium Dioxide*</u> is remarkable example of such non-congruency.**

**New information on the thermophysical properties of phase transitions in Uranium and Uranium dioxide could be particularly valuable for its application in the <u>*nuclear reactor safety*</u> analysis.**

## **PRIORITIES**

- *Uniformity of heating,*
- *Energy deposition control,*
- *Careful choice of the phenomena to be investigated,*

## Seem to be the dominating goals

## **PREFERENCES**

- We should work for *relatively low temperatures* rather than for the highest ones

- We should work for *relatively low densities* rather than for the highest ones

- We should work for the *planar geometry* rather than for other ones.

- We should possibly work for the *defocused beam* rather than for that in-focus

- We should *cut-off the Bragg's peak* rather than to use it for HIB energy deposition.

- We should put the sample *behind* the *beam crossover* rather than to put it just in the focus.

## **MAJOR PREFERENCE**

- When one use the HIB heating in study of thermophysical properties he should strive for its *direct measurement* rather than to withdraw the required information from the results of numerical simulation of final combination of thermophysics, hydrodynamics and heat transfer.

## PHYSICAL PHENOMENA  -

∗ *Critical Point of Gas-Liquid Phase Transition in Single Substances*
   (Location and Properties)

∗ *Critical Point of Gas-Liquid Phase Transition in Complex Substances*
   (Location and Properties)

∗ *Spinodal Decomposition for Gas-Liquid Phase Transition*
   a) Single Substances
   b) Complex Substances

∗ *Metal – Insulator Transition in metals*
   (Location and Properties)

∗ *"Bottom" of "Non-Ideality Valley"*
   (Location and Properties)

****************************

## WHEN STUDYING ALL THE PHENOMENA:   -

> We should work for *relatively low temperatures*
>
> [ $T \sim 0{,}3 \div 2$ eV ]
>
> rather than for the highest ones

> *It is equivalent to:*
>
> $\Delta U \sim 2 \div 10$ kJ/g
>
> *when heating the heavy materials*

****************************

> We should work for *relatively low densities*
>
> $\rho \sim 0{,}1 \div 1{,}0 \, \rho_0$
>
> rather than for the highest ones

> It makes to be expedient
>
> Irradiating of *foam (porous)* samples:
>
> $M \equiv \rho_0/\rho_{00} \sim 3 \div 10$

# PROBLEM OF UNIFORM HEATING

I. Planar geometry

II. Bragg's peak cut-off

III. Putting of the sample BEHIND the beam FOCUS

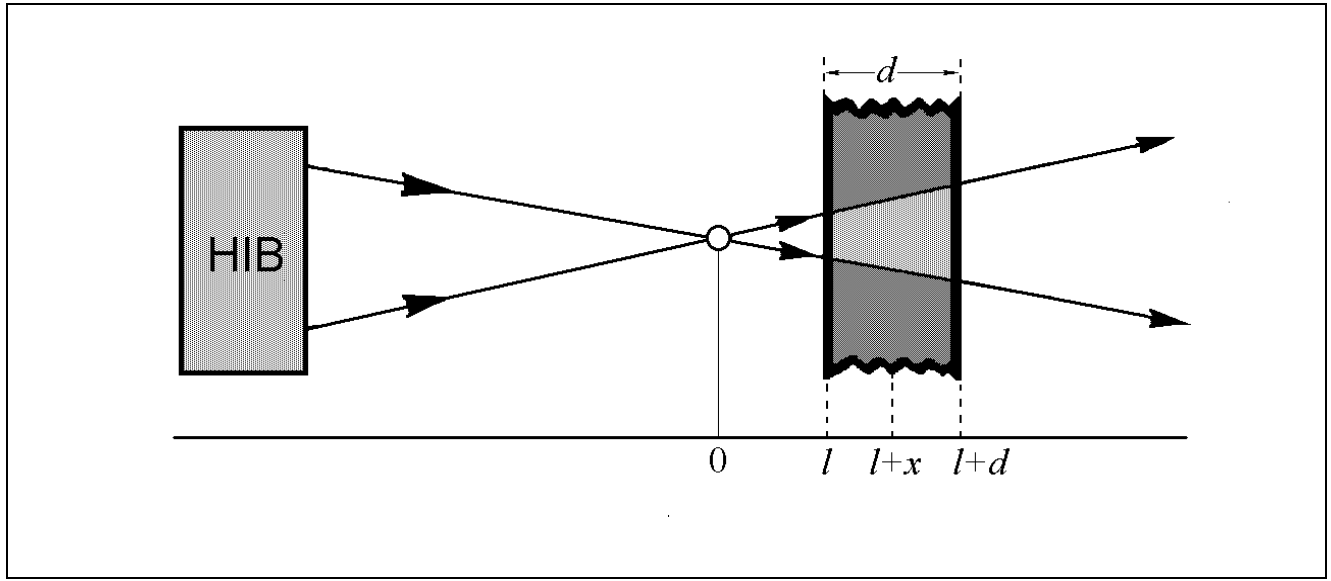

$$\left(\frac{dE}{dx}\right)_{(=HIB)} \sim A(1+Bx)^k \quad \& \quad S(x) = S_0\left(1+\frac{x}{l}\right)^2$$

$$\Downarrow$$

$$\left(\frac{\Delta E}{\Delta V}\right)_{(divergent-HIB)} \sim \frac{1}{S(x)}\left(\frac{dE}{dx}\right)_{(=)} \approx const$$

## HEATING of HIGHLY DISPERSED *POROUS* MATERIALS
⇓

*Quasi-isobaric* heating (!)

## HEATING of *EVAPORATING POROUS* MATERIALS
⇓

Tracing of *Saturation Curve* (!)

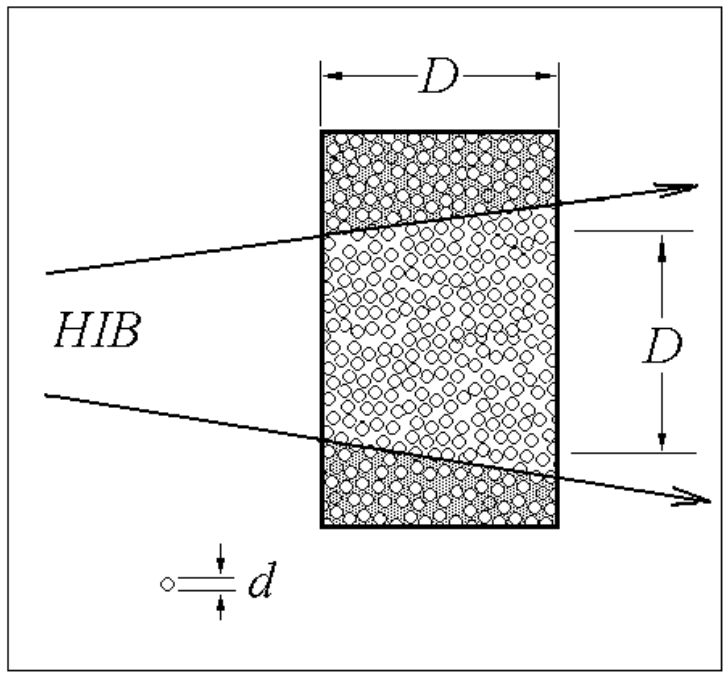

$D \sim 3\ mm \quad d \sim 1\ \mu m \quad \tau_{HIB} \sim 200\ ns \quad a_{(sound)} \sim 5\ km/sec$

⇓

$h \equiv a\, \tau_{HIB} \approx 1\ mm$

⇓

$h \gg d$

*Quasi-free* expansion of every single grain along the saturation curve of material, $P_S(T_S)$, up to the moment, $t^*$, when the total volume of voids will be exhausted by material expansion
($t = t^* \leftrightarrow$ degree of thermal expansion is equal to the porosity ($m \equiv \rho_0/\rho_{00}$))

$\rho_0/\rho(t^*) = m \equiv \rho_0/\rho_{00}$

$t > t^*$
$h < D \leftrightarrow$ quasi-isochoric heating of total sample – *pressure "jump"* (!)

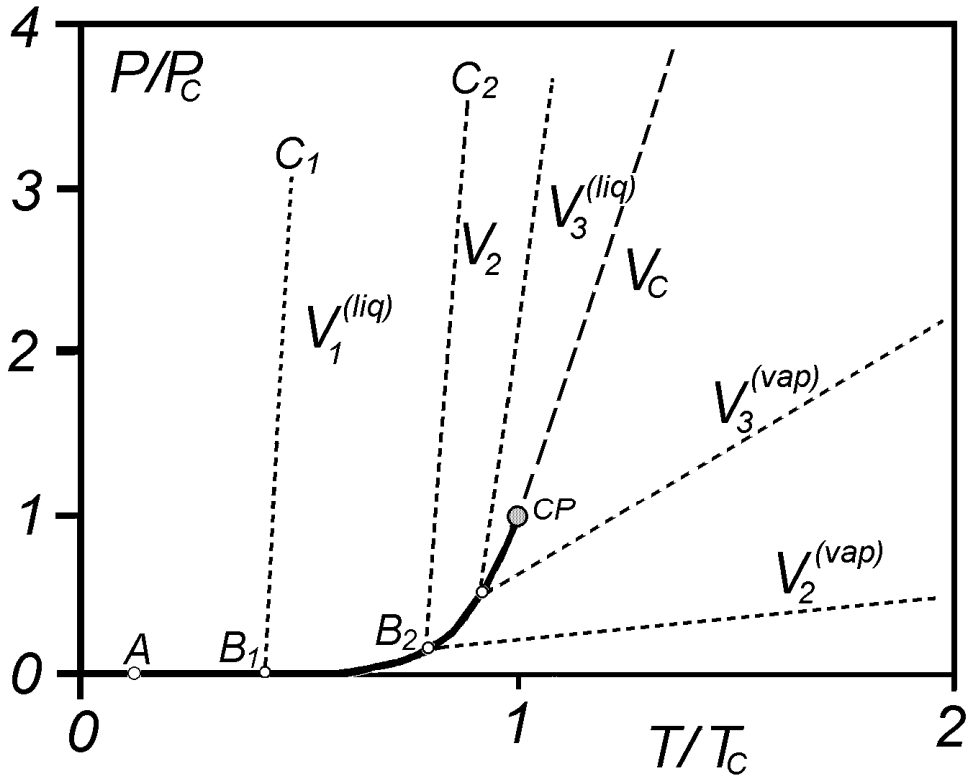

Figure 1. Presumed pressure–temperature trajectories of uniformly heated highly-dispersed evaporating porous material under HIB energy deposition on example of typical pressure–temperature phase diagram in reduced coordinates[1].

*Notations*:
$P_C$ and $T_C$ – critical pressure and temperature; *A-B-CP* – saturation curve of homogeneous material up to the critical point (*CP*); $V_i^{(liq)}$, $V_i^{(vap)}$ – isochors of condensed ("liq") and gaseous ("vap") phases of homogeneous material; $V_C$ – its critical isochor;

*P(T)-Trajectories*: *A* – starting point of heating; $A$-$B_1$-$C_1$ – pressure of heated sample of low porosity ($m \equiv \rho_0/\rho_{00} \sim 2$); $A$-$B_2$-$C_2$ – the same $P(T)$–trajectory for intermediate porosity value ($m \sim 4$); $B_1$, $B_2$ – presumed "pressure jump"; $A$-$B_2$-$V_2^{(vap)}$ – $P(T)$ – trajectory for high porosity value ($m \gg V_C/V_0$); $A$-$B_1$-$B_2$-$CP$-$V_C$ – "critical" $P(T)$–trajectory for special (presumably unknown) porosity value ($m = V_C/V_0$).

---

[1] $P(T)$ – Diagram for One-Component plasma model (from [Iosilevski I. & Chigvintsev A., in "*Physics of Strongly Coupled Plasmas*", Eds. W.Kraeft & M.Schlanges, World Scientific, 1995]).

# MEASUREMENT OF EQUILIBRIUM THERMAL EXPANSION

**IV.** Uniform *isobaric* heating − ( $\Delta t_{heating}$ >> $d/a$ )

**V.** *Planar* geometry

**VI.** Irradiation of sample with *narrow gap* − ( $\Delta d$ << $d$ )

**VII.** Measurements:
- Energy deposition control
- Fixation of the moment of *gap collapse*
  - *Moment of the gap x-ray transparency cut-off*
  - *Moment of the gap conductivity jump*

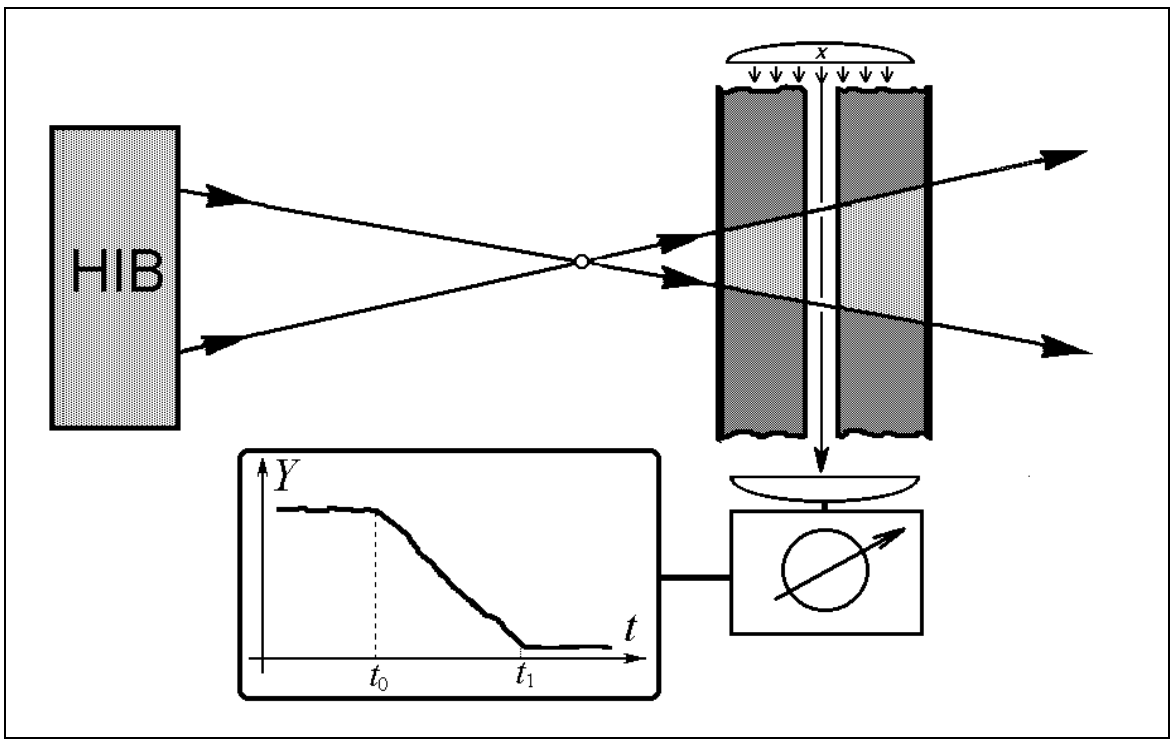

# APPLICATIONS

1) <u>Nuclear safety problem:</u>

   Thermal expansion of Uranium, Uranium dioxide and other nuclear fuels

2) <u>Problem of critical point location:</u>

   *Thermal expansion of anomalous metals* − *U, W, Mo, etc.*

## *Appendix A*:

# An Advanced Equation of State of UO$_2$ up to the Critical Point (*)

*I. Iosilevski* ([+]), *V. Gryaznov* ([++]), *G. Hyland* (**), *C. Ronchi* ([#]), *E.Yakub* ([##]), *V.Fortov* ([++])

An international Project supported by INTAS[2] was started in 1994 with the intent of constructing an equation of state for liquid and gaseous UO$_2$, which fully reproduces the consolidated thermodynamic database of this compound, and, further, provides trustworthy extrapolations up to the critical point.

The equation of state of fluid UO$_2$ was first needed in the analysis of hypothetical reactor accidents where temperatures are expected, at which the vapor pressure of the fuel is sufficiently high to produce liquid mass displacements. In the last decade, however, increasing attention was directed on the effects of the atmospheric environment on the vaporization rate of the molten fuel. On the other hand, empirical thermodynamic properties of urania at high temperatures are presently restricted to the near-stoichiometric or congruent-vaporization compositions. The new equation of state was, therefore, devised for applications encompassing hypo- and hyperstoichiometric compositions [3].

A so-called "Chemical Model" approach [4] was used for the theoretical description of liquid Urania. The model is based on thermodynamic perturbation theory (TPT) modified in order to account for the specific properties of the system studied. It describes, in a unified formalism, a *multi-component mixture* of chemically reactive, strongly interacting neutral and charged molecules and atoms. Their interaction parameters were first expressed theoretically, and then "calibrated" to correctly reproduce sufficiently well established properties of molten UO$_{2.00}$ at the melting point ($T$ = 3120 K).

Comparing the predicted equilibrium vapor pressure with the literature data provided a first validation of the model up to temperatures of the order of 5500 K. A further, positive result is the fairly good agreement of the predicted heat capacity (Fig.1) with the experimental values of Ronchi *et al.* [5], which extend up to 8000 K. Up to now, none of the existing liquid UO$_2$ models was able to correctly reproduce the observed dependence of $C_p$ on temperature. The calculations were finally extrapolated to the high-pressure and high-temperature region ($P \leq$ 1 GPa, $T \leq$ 20000 K).

A relevant result is the unusual structure of the gas-liquid phase transition boundary (Figs.2 and.3), which exhibits a *striking difference* with respect to that of ordinary liquids. This feature is caused by the *non-congruency* of the coexisting phases in UO$_{2 \pm x}$. Both the total vapor pressure and the degree of oxygen enrichment of the vapor phase strongly depend on the evaporation regime. Two distinct pressure-temperature functions $P=P(T)$ were, therefore, calculated (Fig.2), corresponding to the limiting cases of: a) *boiling,* i.e. *slow* evaporation under "global equilibrium conditions", and b) *saturation,* i.e. *fast,* non-equilibrium evaporation (the so-called "forced-congruent mode").

A characteristic feature of non-congruent vaporization in UO$_{2 \pm x}$ is the production of a very high maximal vapor pressure ($P_{max}$ ~ 1 GPa) as well as a substantial oxygen enrichment of the vapor phase over boiling UO$_2$ ((O/U)$_{max}$ ~ 7), as shown in Figs. 2 and 3. The implicit correlation of the total fuel vapor pressure with the effective oxygen chemical potential is obviously of great importance for a correct prediction of the fuel vaporization rate in a real reactor accident scenario.

The critical point of *truly non-congruent* phase transition in UO$_2$ was also calculated. This point *essentially differs* from that defined for a gas-liquid phase transition in *simple liquids;* in particular, we have here $(\partial P/\partial V)_T \sim (P/V) \neq 0$. The predicted critical parameters are: $T_c \approx$ 10120 K, $P_c \approx$ 965 MPa, $\rho_c \approx$ 2.61 gcm$^{-3}$.

* * * * * * * * * * * * * * * * * * * * * * * * * * * * * * * * *

[+]    - Moscow Institute of Physics and Technology, [StateUniversity] (Russia)
[++]   - Institute of Chemical Physics in Chernogolovka (Russia)
**     - University of Warwick, Coventry (United Kingdom)
[#]    - European Commission, JRC, Institute for Transuranium Elements, Karlsruhe (Germany)
[##]   - Odessa Medical State University  (Ukraine)

* * * * * * * * * * * * * * * * * * * * * * * * * * * * * * * * *

(*)  Iosilevski I.L., Hyland G.J. Ronchi C. and Yakub E.S., *Trans. Amer. Nucl. Soc*. **81** 122 (1999)

---

[2] International Association for the Promotion of Cooperation with Scientists from the Independent States of the Former Soviet Union.
[3] Final Report of Project INTAS-93-0066, European Commission, JRC-ITU, Karlsruhe, Germany (1997).
[4] V. Gryaznov, I. Iosilevski, V. Fortov *et al.,* "Thermophysical Properties of Working Fluids of Gas-Core Nuclear Reactor", /Ed. V.Iewlev (1980),  ATOMIZDAT, Moscow (in Russian).
[5] C. Ronchi, J-P Hiernaut, R. Selfslag, G.J. Hyland,  *Nucl. Sci.and Eng.* **113**, 1 (1993).

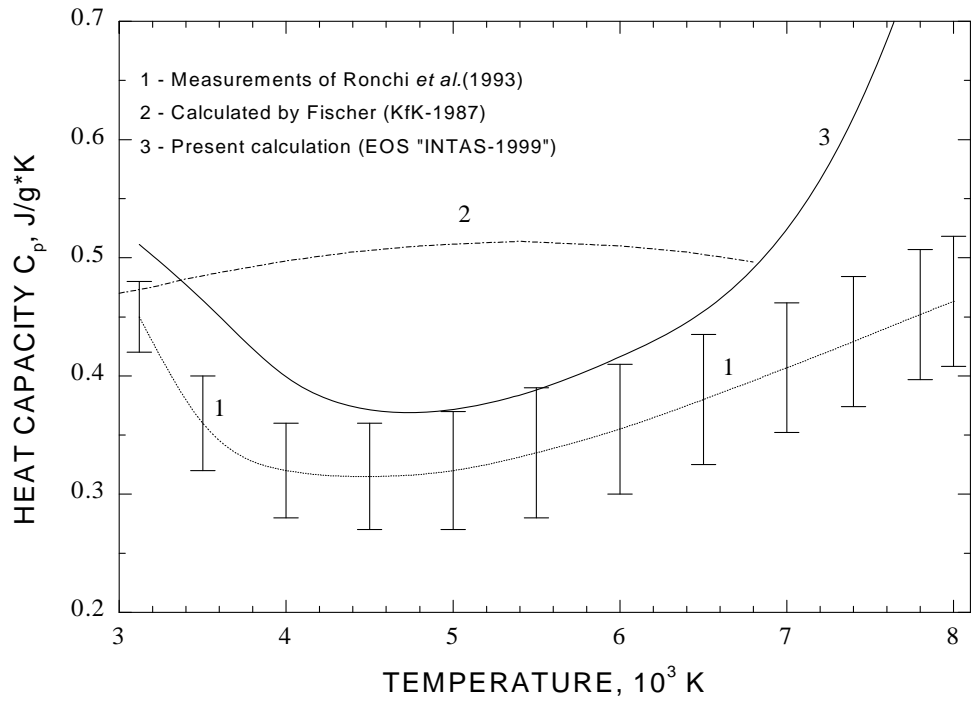

FIG.1. Comparison of predicted and observed heat capacities of liquid uranium dioxide

*1* - Measurements of Ronchi *et al.* (1993); *2* - Calculated Cp of Fischer (KfK-1987);
*3* - Present calculation (INTAS-99 EOS)

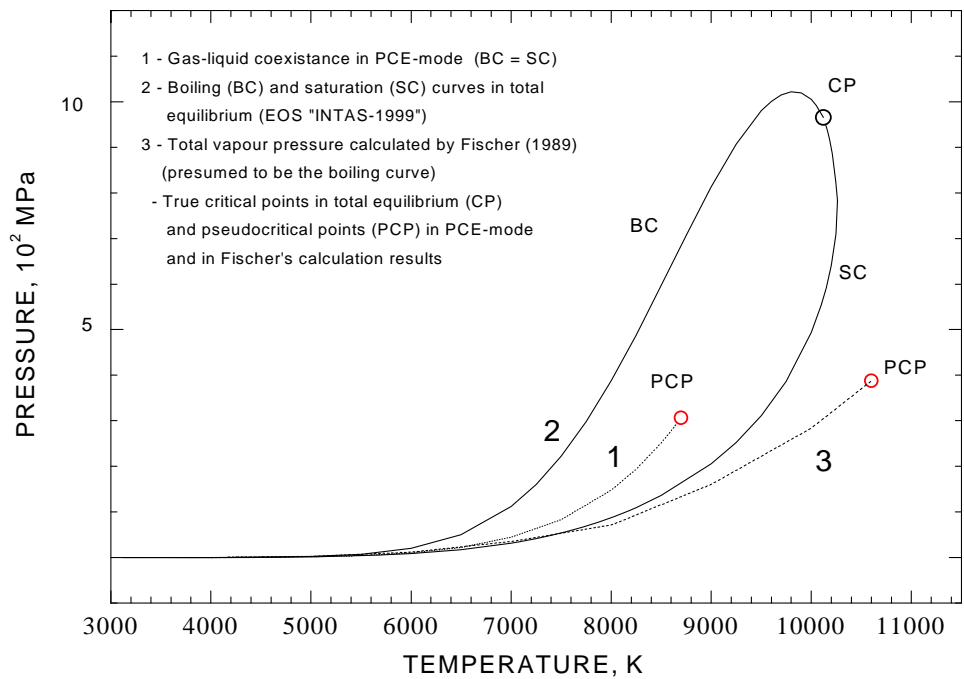

FIG.2. Calculated gas-liquid (P-T) coexistance in UO$_2$.

*1* - Gas-liquid coexistance in PCE-mode (BC = SC)
*2* - Boiling (BC) and saturation (SC) curves in total equilibrium (INTAS-99 EOS)
*3* - Total vapour pressure calculated by Fischer (1989)
    (presumed to be the boiling curve)
*CP, PCP* - True critical points in total equilibrium (CP) and pseudocritical points (PCP)
    in PCE-mode and in Fisher's calculation results

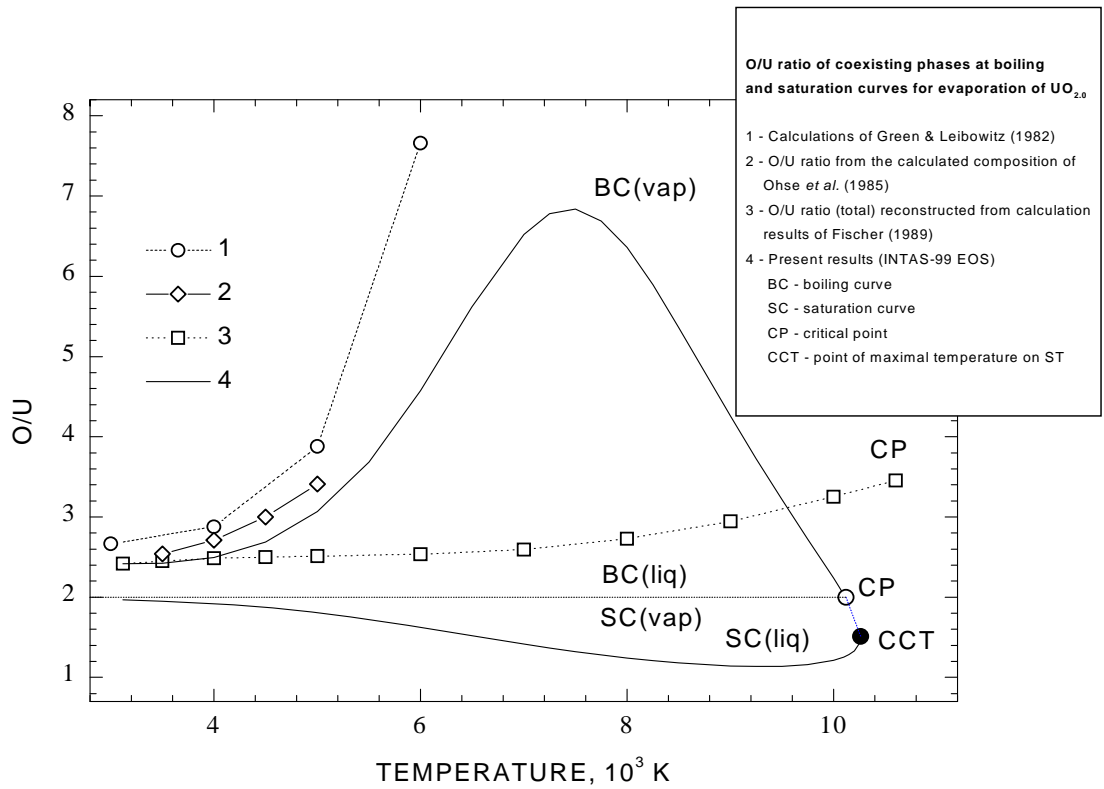

Figure 3.
O/U ratio of coexisting phases at boiling and saturation curves for evaporation of $UO_{2.0}$

    *1* - Calculation results of Green & Leibovitz (1982)
    *2* - O/U ratio from the calculated composition of Ohse *et al.* (1985)
    *3* - O/U ratio (total) reconstructed from calculation results of Fischer (1989)
    *4* - Present results (INTAS-99 EOS)
      BC - boiling curve
      SC - saturation curve
      CP - critical point
     CCT - point of maximal temperature on SC